\documentclass[journal,article,submit,pdftex,moreauthors]{Definitions/mdpi} 

\firstpage{1} 
\pubvolume{1}
\issuenum{1}
\articlenumber{0}
\pubyear{2025}
\copyrightyear{2024}
\datereceived{ } 
\daterevised{ }
\dateaccepted{ } 
\datepublished{ } 
\usepackage{braket} 
\hreflink{https://doi.org/} % If needed use \linebreak

\Title{Effect of the Coulomb interaction on nuclear deformation and drip lines}

\TitleCitation{Effect of the Coulomb interaction on nuclear deformation and drip lines}

\Author{Kenta Hagihara $^{1,\dagger}$*,
Takashi Nakatsukasa $^{2,3,4\dagger}$*\orcidA{},
and Nobuo Hinohara $^{2,3\dagger}$\orcidB{}}

\AuthorNames{Kenta Hagihara, Takashi Nakatsukasa, Nobuo Hinohara}

\isAPAStyle{%
       \AuthorCitation{Hagihara, K., Nakatsukasa, T., \& Hinohara, N.}
         }{%
        \isChicagoStyle{%
        \AuthorCitation{Hagihara, Kenta, Takashi Nakatsukasa, and Nobuo Hinohara.}
        }{
        \AuthorCitation{Hagihara, K.; Nakatsukasa, T.; Hinohara, N.}
        }
}

\address{
$^{1}$ \quad Graduate School of Science and Technology,
University of Tsukuba, Tsukuba 305-8571, Japan\\
$^{2}$ \quad Center for Computational Sciences, University of Tsukuba,
University of Tsukuba, Tsukuba 305-8577, Japan\\
$^{3}$ \quad Faculty of Pure and Applied Sciences, University of Tsukuba,
University of Tsukuba, Tsukuba 305-8571, Japan\\
$^{4}$ \quad RIKEN Nishina Center, Wako 351-0198, Japan\\
}
\corres{Correspondence:
hagihara@nucl.ph.tsukuba.ac.jp;
nakatsukasa@nucl.ph.tsukuba.ac.jp (T.N.)
}
\firstnote{These authors contributed equally to this work.}

\abstract{
Nuclei are self-bound systems in which the strong interaction
(nuclear force) plays a dominant role
and the isospin is approximately a good quantum number.
The isospin symmetry is primarily violated by the electromagnetic
interactions, namely the Coulomb interaction among protons,
effects of which need be studied to understand
importance of the isospin symmetry.
We investigate the effect of the Coulomb interaction on nuclear properties,
especially the quadrupole deformation and neutron drip line,
utilizing the density functional method which provides
a universal description of nuclear systems in the entire nuclear chart.
We carry out calculations of even-even nuclei
with the proton number $2\leq Z \leq 60$.
The results show that the Coulomb interaction plays a significant role in enhancing the quadrupole deformation across a wide range of nuclei. 
We also find that nuclei near the neutron drip line
gain an additional binding energy by the Coulomb interaction,
which may lead to a shift of the neutron drip line
toward a larger neutron number.
}


\begin{document}
%%%%%%%%%%%%%%%%%%%%%%%%%%%%%%%%%%%%%%%%%%
\section{Introduction}
%大まかな流れ⇨関連する先行研究⇨先行研究を踏まえた自分の研究

%\subsection{Nuclei as many-body system}
\subsection{Basic properties of nuclei}
%general properties of nuclei and the mothods for them, for the concerence of "many-body"

Nuclei are composed of two types of fermions, neutrons and protons,
which are often regarded as identical particles (called ``nucleons'')
with different isospins $T_z$.
The number of nucleons (the sum of the proton and neutron numbers)
is the mass number $A$.
The reason why large nuclei with $A>300$ do not exist in nature
is due to the repulsive Coulomb interaction among protons,
which is the main subject of the present paper.

There are well-known properties of nuclei, such as
the saturation density of nuclei $\rho_0\approx 0.16$ fm$^{-3}$), 
constant binding energy per nucleon ($B/A\approx 8$ MeV),
nucleon's mean free path which is larger than the nuclear size,
and the magic number of neutrons and protons (2, 8, 20, 28, 50, 82, $\cdots$)
\cite{BM69,BM75}.
Nuclear shapes are also of great interest,
and have been studied extensively in the past
\cite{AFN90}.
The collective motion associated with the shape degrees of freedom
may appear at low energy,
which is due to a nuclear property of the approximately constant $B/A$;
a nucleus can be divided into two fragments without the cost of much energy.
Thus, it is relatively free to be transformed into different shapes.
Nuclear deformation is an example of the spontaneous breaking of
the rotational symmetry in finite quantum many-body systems
\cite{Ald56,NMMS16,NMMY16}.
A variety of nuclear shapes are present in nuclei, including
octupole shapes that violate the parity symmetry as well.
Among these, the axially symmetric quadrupole deformation
is the most common shape in nuclei \cite{BM75},
which will be discussed in the following sections.

\subsection{Energy density functional method}

It is still difficult even now to directly solve
the Schr\"odinger equation for finite nuclei with many nucleons.
This is partially due to singular, complex, and many-body characters
of the nuclear force,
and also to the treatment of the center-of-mass motion
for self-bound systems.
Among many other alternative methods,
the energy density functional (EDF) theory,
which we use in this study,
is one of the most powerful tools.
In nuclear physics, it is often identified with the mean-field theory
for density-dependent interactions,
in which
each nucleon independently moves in an average potential
that is produced by the interactions among nucleons themselves.
In contrast to the electrons in atoms and molecules,
there is no external potential to bind the nucleons.
The nucleons are bound by a potential created by the nucleons themselves.

% Interestingly, in nuclear physics, there is a fact that the binding energy per nucleon except for very light nuclei is almost constant and the value is approximately $8 MeV$, 
% which tells us that each nucleon interacts with nearest neighbours and the number of interactions is not proportional to the number of nucleon pairs $A(A-1)$, but simply to A.
% In addition, nucleon mean free path has been found to be larger than the radius of the nuclei.
% These phenomena let us think that nucleon in nuclei can be treated in terms of nucleon moving independently in an average potential, 
% despite of the nuclear force which works between all pairs of two nucleons.
% But still, it is so difficult to calculate all particles interacting each other that many kinds of alternative method have been developed, 
% and the energy density functional theory (EDFT), which we used in this study, is one of the powerful tools.
% This framework was firstly proposed by P. Hohenberg and W. Kohn in 1964.
% They indicated that for a system of fermions that are subject to an external potential $v_{ext}$, 
% the total energy can be written as a functional of the particle density $\rho(\boldsymbol{r})$ which is composed of the sum of single-particle orbitals $\phi _j(\boldsymbol{r})$.

\subsection{Electromagnetic interaction in nuclei}
The electromagnetic force plays a crucial role
for many nuclear phenomena.
For instance, the alpha decay is described by a quantum tunneling
in which an alpha particle travels through the Coulomb barrier
potential \cite{Gam28}.
It is also responsible for
the fact that there exist only a finite number of species of elements
on Earth.
The repulsive Coulomb force among protons disfavors
nuclei with large $Z$,
which eventually leads to instability against nuclear fission
\cite{BW39}.
$^{238}$U is known as the heaviest ``stable'' nucleus on Earth.
In fact, $^{238}$U is unstable with respect to the alpha decay,
however, its lifetime is approximately the same as the age of the Earth
(4.5 billion years),
thus, it is normally categorized as a stable nucleus.

The liquid drop model
is one of the most successful nuclear models
to represent the saturation property.
It is based on the classical liquid drop picture
that always favors the spherical shape.
Therefore, the deformed shape in the ground state of a nucleus
is a consequence of the quantum mechanical effect,
which is often referred to as ``shell energy'' or ``shell correction''
\cite{Bra72}.
However, one may expect that the magnitude of the deformation is
influenced by the Coulomb interaction,
because the Coulomb energy favors a non-spherical shape \cite{BM69,RS80}.

Another prominent electromagnetic effect can be found in the
position of the proton drip line.
The liquid drop model has a symmetry-energy term that favors
nuclei with equal numbers of neutrons and protons, $N=Z$.
However, there are no stable $N=Z$ nuclei beyond $^{40}$Ca.
If the isospin symmetry is exact,
the nuclear chart should be symmetric with respect to the $N=Z$ line.
In reality, the $N-Z$ symmetry of the nuclear chart is significantly violated,
and the proton drip line is located at $N>Z$ for heavy nuclei with $A>100$.
This is a trivial effect on the proton drip line.
In this paper,
we show a possible nontrivial Coulomb effect on the neutron drip line.

\subsection{Scope of the present paper}

We perform numerical calculations based on
the Skyrme EDF.
In addition,
we perform calculations neglecting the Coulomb interaction.
This makes it possible to estimate the effect of the Coulomb interaction
on the nuclear properties.
In this paper, we focus our discussion on changes
in the quadrupole deformation and
position of the neutron drip line.

%%%%%%%%%%%%%%%%%%%%%%%%%%%%%%%%%%%%%%%%%%
\section{Hartree-Fock-Bogoliubov theory}
The Hartree-Fock-Bogoliubov (HFB) theory
is a powerful tool to treat many-body systems
in which the mean-field effects are dominant in both the particle-hole and
particle-particle (hole-hole) channels.
The HFB equation, derived by minimizing the EDF $E[\rho,\kappa]$,
where $\rho$ and $\kappa$ are the particle and pair densities,
respectively,
can be solved by diagonalizing the HFB Hamiltonian.
The solutions give the Bogoliubov transformation
to define quasi-particles $(\alpha_k,\alpha_k^\dagger)$
and their energies $E_k$.
The quasi-particle energies have a gap, $E_k>\Delta$,
when the system is in the superfluid phase.

%Generally, two-body Hamiltonian can be expressed in terms of the creation $\hat{c}^{\dagger} _i$ and the annihilation $\hat{c} _i$ operators with an anti-symmetrized two-body interaction matrix elements $\bar{v} _{ijkl} = \bra{ij} V \ket{kl -lk}$.
%\begin{align}
%    \hat{H} = \sum _{i j} e_{i j} \hat{c}^{\dagger} _{i} \hat{c} _{j} + \frac{1}{4} \sum _{i j k l} \bar{v}_{i j k l} \hat{c}^{\dagger}_{i} \hat{c}^{\dagger}_{j} \hat{c}_{l} \hat{c}_{k} 
%\end{align}
Starting from a nuclear EDF $E[\rho,\kappa]$,
the HFB equation is given in the form of
the eigenvalue equation.
\begin{equation}
\begin{pmatrix}
h-\lambda  & \Delta \\
-\Delta^* & -(h-\lambda)^*
\end{pmatrix}
\begin{pmatrix}
U_k \\
V_k
\end{pmatrix}
= E_k
\begin{pmatrix}
U_k \\
V_k
\end{pmatrix}
,
\label{eq:HFB_equation}
\end{equation}
where
\begin{equation}
h_{nn'}\equiv\frac{\delta E}{\delta\rho_{n'n}},
\quad \quad
\Delta_{nn'}\equiv\frac{\delta E}{\delta\kappa^*_{nn'}},
\label{eq:h_and_Delta}
\end{equation}
and $\lambda$ is the Fermi energy (chemical potential).
In the HFB method, the wave function of the ground state $\ket{\phi}$ is defined as the quasi-particle vacuum $\alpha_{k} \ket{\phi} = 0$, 
where the quasi-particle creation and annihilation operators
($\alpha$, $\alpha^{\dagger}$)
are connected to the particle creation and annihilation operators ($c$, $c^{\dagger}$)
via a unitary transformation, namely the Bogoliubov transformation
% \begin{align}
%     \alpha_{k}=\sum_{n} (U_{nk}^* c_n + V_{nk}^* c_n ^{\dagger}), \quad 
%     \alpha_{k} ^{\dagger}=\sum_{n} (U_{nk} c_n ^ {\dagger} + V_{nk} c_n ^{\dagger}), 
%     \quad
%     \begin{pmatrix}
%         \alpha \\
%         \alpha^{\dagger}
%     \end{pmatrix}
%     =
%     \begin{pmatrix}
%         U^{\dagger} & V^{\dagger} \\
%         V^{T} & U^{T}
%     \end{pmatrix}
%     \begin{pmatrix}
%         c \\
%         c^{\dagger}
%     \end{pmatrix}.
% \end{align}
\begin{align}
    \alpha_{k}=\sum_{n} (U_{nk}^* c_n + V_{nk}^* c_n ^{\dagger}), \quad 
    \alpha_{k} ^{\dagger}=\sum_{n} (U_{nk} c_n ^ {\dagger} + V_{nk} c_n ^{\dagger}). 
    % \quad
    % \begin{pmatrix}
    %     \alpha \\
    %     \alpha^{\dagger}
    % \end{pmatrix}
    % =
    % \begin{pmatrix}
    %     U^{\dagger} & V^{\dagger} \\
    %     V^{T} & U^{T}
    % \end{pmatrix}
    % \begin{pmatrix}
    %     c \\
    %     c^{\dagger}
    % \end{pmatrix}.
\end{align}
In terms of the quasi-particle wave functions $(U,V)$,
the one-body density matrix $\rho$ and the pair density $\kappa$
are given as
\begin{align}
    \rho_{n n^{\prime}} = \bra{\phi} c_{n^{\prime}}^\dagger c_n \ket{\phi} = (V^* V^T)_{nn^{\prime}}, \quad 
    \kappa_{n n^{\prime }} = \bra{\phi} c_{n^{\prime}} c_n \ket{\phi} = (V^* U^T)_{nn^{\prime}}.
\end{align}
Since the HFB Hamiltonian, the matrix on the left-hand side of
Eq. \eqref{eq:HFB_equation},
depends on the densities, $\rho$ and $\kappa$,
we iteratively solve the equation
until the self-consistency is achieved.

In the present paper, we discuss the axial quadrupole deformation
which is characterized by the following deformation parameter
\begin{align}
    \beta_2 &= \frac{\sqrt{\pi}}{5} \frac{\langle \hat{Q}_2 \rangle}{A \langle r^2 \rangle} \\
    \hat{Q}_l &= r^l Y_{l0}(\theta,\phi) =\sqrt{\frac{2l+1}{4\pi}} r^l P_{l}(\cos\theta) .
\end{align}

The quasi-particle states and energies characterize the excitation properties
of the superfluid system \cite{RS80}.
For a numerical purpose and a physical interpretation,
it is sometimes convenient to define the canonical single-particle 
states and their energies.
The canonical single-particle states $d_i^\dagger$ are defined as
those that diagonalize the density matrix $\rho$.
\begin{equation}
d_i^\dagger=\sum_n \phi_{ni} c_n^\dagger \ ,
\quad\quad
\sum_{nn'} \rho_{nn'}\phi_{n'i}=v_i^2 \phi_{ni} .
\end{equation}
The eigenvalues $v_i^2\geq 0$ correspond to the occupation
number in the Bardeen-Cooper-Schrieffer (BCS) theory,
though the canonical states $i$ are not the energy eigenstates.
The canonical single-particle energies $e_i$ are defined by
the expectation value of $h$ in Eq.~\eqref{eq:h_and_Delta};
$e_i\equiv \braket{\phi_i|h|\phi_i}
=\sum_{nn'} \phi_{ni}^* h_{nn'} \phi_{n'i}$.

% hfbtho_v3.0
% hfodd
% \begin{align}
%     \hat{Q}_{\lambda 0}(r,\theta)= \mathcal{N}_{\lambda} \sqrt{\frac{2\lambda+1}{4\pi}} r^{\lambda} P_{\lambda} (\cos\theta).\\
% \end{align}
% \begin{align}
%     \mathcal{N}_{0} &  =\sqrt{4\pi} \\
%     \mathcal{N}_{1} &  =\sqrt{\frac{4\pi}{3}}\frac{1}{10}\\
%     \mathcal{N}_{2} &  =\sqrt{\frac{16\pi}{5}}\frac{1}{100} \\
%     \mathcal{N}_{\lambda} &  =\frac{1}{10^{\lambda}},\ \ \lambda>2
% \end{align}

%%%%%%%%%%%%%%%%%%%%%%%%%%%%%%%%%%%%%%%%%%
\section{Results}

The numerical calculation is performed using
an open source code {\sc hfbtho} \cite{HFBTHO17}.
We carry out a systematic calculation for even-even nuclei with $Z=2-60$
using the SLy4 parameter set of the Skyrme functional \cite{CBHMS97}
with a mixed-type pairing. 
The calculation was performed by solving
the self-consistent HFB equation
using the harmonic oscillator basis.
The initial states of the iteration are set
to have seven different quadrupole deformations
with $\beta_2=-0.3, -0.2, -0.1, \cdots$, and $0.3$.
At the beginning of the self-consistent procedure,
we perform the HFB calculation with
a constraint on the quadrupole deformation.
After ten iterations,
we release the constraint
to find an optimal deformation.
Comparing the converged energies with the different initial states,
we find the ground state with the minimum energy.
Using the ground-state wave function,
we obtain quantities
such as total energy, quadrupole deformation,
canonical single-particle energies.
Then, we repeat the same calculation
neglecting the Coulomb interaction among protons.

\subsection{Quadrupole deformation}

\begin{figure}[H]
\begin{center}
\includegraphics[width=130mm]{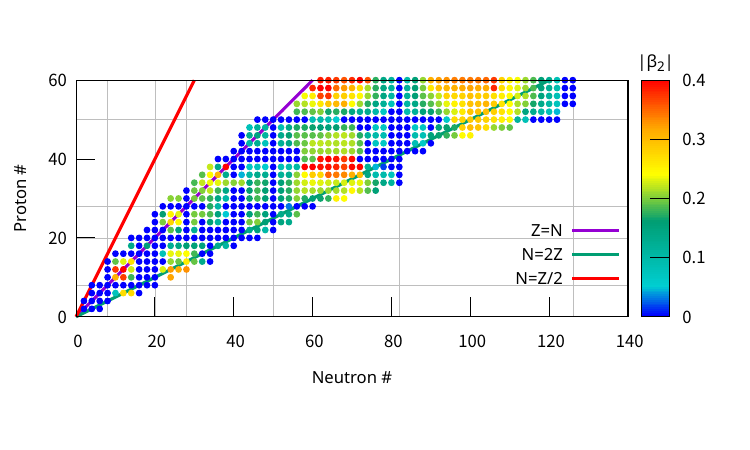}
\caption{Deformation parameter $|\beta_2|$ of the calculated ground states
of even-even nuclei with the SLy4 EDF including the Coulomb EDF.
Those with positive two-nucleon separation energy (negative Fermi energy)
are shown.
The prolate and oblate shapes are not distinguished.
The lines of $N=Z$, $N=2Z$, and $Z=2N$ are shown
for reference.
}
\label{fig:abs_beta2_C}
\end{center}
\end{figure}

\begin{figure}[H]
\begin{center}
\includegraphics[width=130mm]{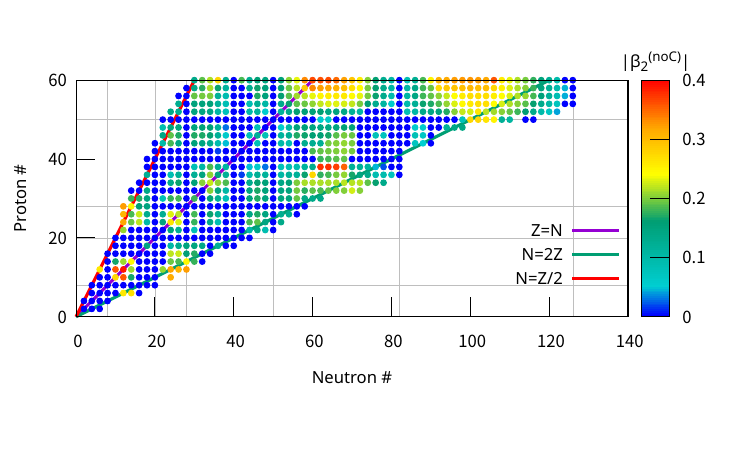}
\caption{
Same as Figure~\ref{fig:abs_beta2_C} but calculated
without the Coulomb interaction,
$|\beta_2^\text{(noC)}|$.
}
\label{fig:abs_beta2_NC}
\end{center}
\end{figure}

%general talk of the deformation
We show the results for the calculated quadrupole deformation $\beta_2$
for the cases where the Coulomb interaction is present and absent,
in Figures \ref{fig:abs_beta2_C} and \ref{fig:abs_beta2_NC},
respectively.
% In the plot, we show nuclei with negative Fermi energies
% both for protons and for neutrons, $\lambda_{n(p)}<0$.
% Those with the positive $\lambda$'s indicate
% that they are particle unbound.
In the plot, we show nuclei with positive two-neutron-separation energy $S_{2n}(Z,N) = E(Z,N+2)-E(Z,N) > 0$. 
Those with the positive separation energy indicates that they are unbound particles.

Although nuclei in the proton-rich side become unbound
because of the Coulomb energy,
for most nuclei,
the spherical (deformed) nuclei 
stays spherical (deformed)
regardless of the presence of the Coulomb interaction.
This confirms our understanding of the nuclear deformation:
The classical liquid drop always favors spherical,
but the quantum shell effects
drive the nucleus to be deformed.
Since the shell structure is mainly determined by
the nuclear force (strong interaction),
we can expect that the electromagnetic effects are minor in the
shell effects.

% In figure \ref{fig:diff}, the color variation represents the degree of deformation change in the nucleus due to the Coulomb interaction.  The blue indicates a decrease in deformation, while the purple indicates that there is no change in deformation. The red represents nuclei that exhibit an increase in deformation due to the Coulomb interaction. 
% \begin{figure}[H]
% \begin{center}
% \includegraphics[width=100mm]{./PDF_RPMBT/diff_abs_beta.pdf}
% \caption{The change of $\beta_2$ deformation between two situations}
% \label{fig:diff}
% \end{center}
% \end{figure}

% change in the deformation

\begin{figure}[bh]
\begin{center}
\includegraphics[width=130mm]{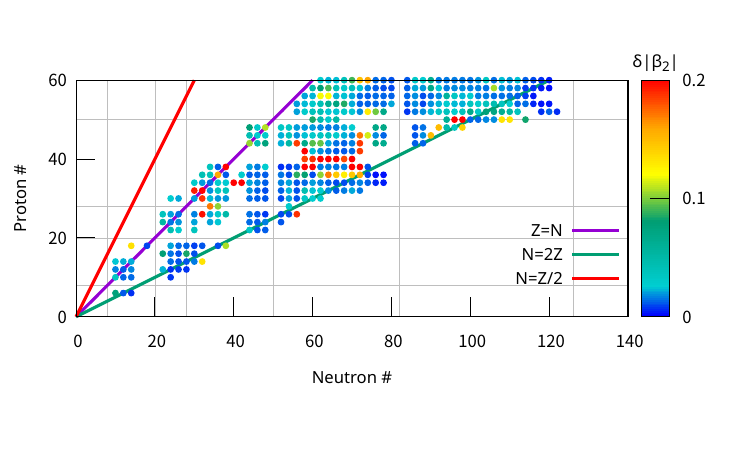}
\caption{Difference in deformation $|\beta_2|$ between those calculated
with and without the Coulomb interaction;
$\delta|\beta_2|\equiv |\beta_2| - |\beta_2^\text{(noC)}|$
The blank spaces indicate the spherical nuclei
with $\delta|\beta_2|=0$.
}
\label{fig:how_diff}
\end{center}
\end{figure}

As mentioned earlier, the Coulomb interaction plays a minor role
in the determination of whether the nucleus is deformed or spherical.
However, it influences the magnitude of the deformation.
As a general trend, the Coulomb interaction enhances the 
quadrupole deformation.
In Figure \ref{fig:how_diff}, we show the difference in $|\beta_2|$
between those with and without the Coulomb interaction.
The blank spots are spherical nuclei which keeps the sphericity
in both calculations.
A significant change in the deformation is caused by
the shape coexistence (multiple potential-energy minima).
The inclusion of the Coulomb interaction changes the ordering in energy
of the potential local minima, which leads to a jump in the location of
absolute minimum.
This is seen in many of Zr isotopes ($Z=40$) with $N=60-70$.
It should be noted that
the superposition of different shapes is often required
for such cases with approximately degenerate multiple potential minima.

\subsection{Neutron drip line}
The electromagnetic interaction makes many nuclei unstable against the proton emission.
This is clearly seen in Figures~\ref{fig:abs_beta2_C} and \ref{fig:abs_beta2_NC}.
A large number of nuclei in the proton-rich side of Figure~\ref{fig:abs_beta2_NC}
are absent in Figure~\ref{fig:abs_beta2_C},
becoming unbound with respect to the (two-)proton emission.
Since the Coulomb barrier may hold unbound protons for a while,
some of them may be metastable proton emitters.
In addition, the Coulomb interaction makes heavy nuclei unstable against the fission
and the alpha decay.
These are well-known and naively expected effects of the electromagnetic interaction
for nuclei.

In contrast, the results seen on the neutron-rich side are somewhat unexpected.
The repulsive Coulomb interaction provides an additional binding effect to
neutron-rich nuclei near the drip line.
Some unbound nuclei calculated with only the nuclear part of the EDF
become bound by including the Coulomb interaction among protons.
The Coulomb interaction between two protons has a repulsive nature,
thus, the total binding energy is always reduced.
However, the position of the neutron drip line is determined by
the two-neutron separation energy which is
the difference in the binding energy between $N$ and $N-2$ neutron systems
\footnote{The odd-even mass difference in nuclei predicts the mass difference
between even-$N$ and odd-$N$ of order of 1 MeV,
which is significantly larger than the calculated two-neutron separation energy
near the drip line.
}.
The two-neutron separation energy is well approximated by
$-2\lambda_n$ where $\lambda_n$ is the neutron Fermi energy
in Eq.~\eqref{eq:HFB_equation}.

\begin{figure}[t]
\begin{center}
\includegraphics[width=100mm]{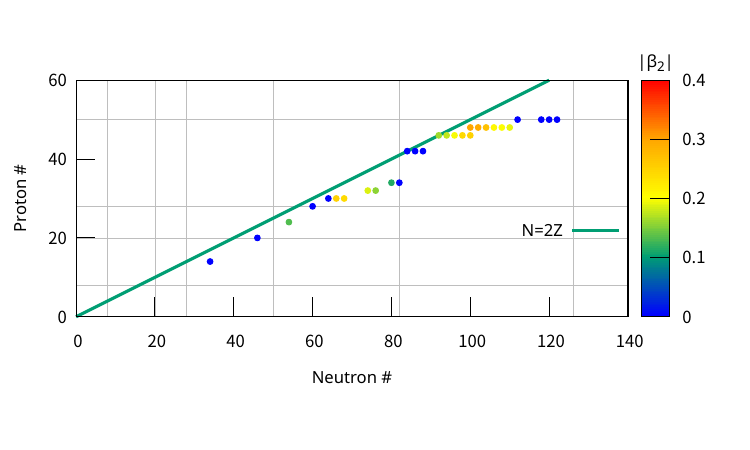}
\caption{Nuclei bound by the inclusion of the Coulomb interaction.
In other words,
those exist in Figure \ref{fig:abs_beta2_C}
but are missing in Figure \ref{fig:abs_beta2_NC}.
The color indicates $|\beta_2|$.
}
\label{fig:bound_nuclei}
\end{center}
\end{figure}

Figure~\ref{fig:bound_nuclei} shows that
29 bound nuclei in Figure~\ref{fig:abs_beta2_C} 
are actually unbound in Figure~\ref{fig:abs_beta2_NC}
without the Coulomb interaction.
We see that these contain both spherical and deformed nuclei,
which means that
it is not due to a gain in the Coulomb energy caused by the deformation.

\subsection{Canonical single-particle energy of neutrons}

\begin{figure}[t]
\begin{center}
\includegraphics[width=100mm]{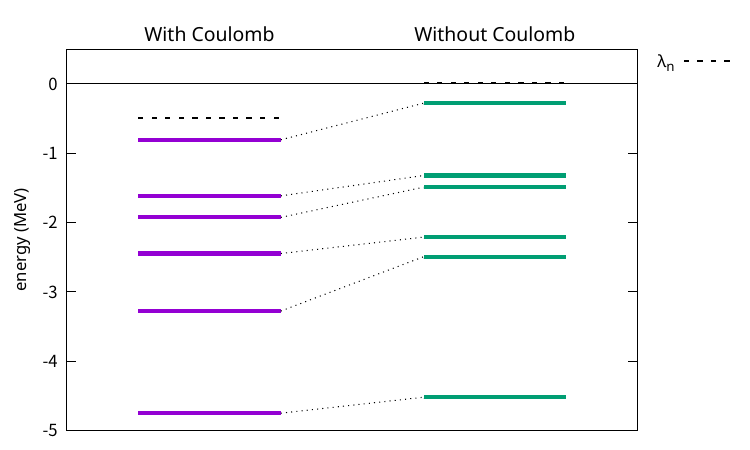}
\caption{Canonical single-particle energies for neutrons in $^{170}$Sn,
including and excluding the Coulomb interaction in the left (magenta line)
and right panel (green line), respectively.
The orbits are, from the top to the bottom, $1i_{13/2}$, $3p_{1/2}$, $2f_{5/2}$,
$3p_{3/2}$, $1h_{9/2}$, and $2f_{7/2}$.
The neutron Fermi energy (chemical potential) is
shown by the black dashed line.
}
\label{fig:levels}
\end{center}
\end{figure}

To examine the mechanism of the shift of the neutron drip line, we investigate the canonical single-particle energies $e_i$ of neutrons.
Figure \ref{fig:levels} shows them for $^{170}$Sn
near the threshold $-5\mbox{ MeV}<e_i<0.5$ MeV.
This nucleus has a spherical shape and is one of the nuclei
that are bound only when the Coulomb interaction is included.
The inclusion of the Coulomb interaction
lowers the energies by several hundreds of MeV,
which affects the Fermi energy $\lambda_n$ leading to the increase of the
neutron separation energy.

The Coulomb interaction cannot increase the depth of the attractive potential for neutrons.
We think the change in $\lambda_n$ cannot be explained
in the scope of the classical mechanics.
It is a quantum mechanical effect to shift the neutron energies
through a change in the density distribution.
Due to the repulsive character of the Coulomb interaction,
the nuclear radius slightly increases.
The increase in the radius decreases the potential depth
and expands the potential range.
For single-particle energies near the bottom of the potential,
the former effect is dominant, and the single-particle energies increase.
In contrast, near the threshold $e_i\approx 0$,
the latter effect dominates and
the kinetic energy decreases because of
the increased spread of the single-particle wave functions.
This is a quantum mechanical effect associated with
the uncertainty principle for the position and momentum.
The relationship between the expansion of the region and the energy can be explained in a framework of the quantum mechanics. 

% One-particle in one-dimensional infinite well potential has the eigen energy given by the following form.
% \begin{align}
% \ E_{n} = n^2 (\pi^2 \hbar^2 /2m) L^{-2}
% \end{align}
% It is obvious that if the region where the existence of a particle is expanded, the energy becomes smaller.

%%%%%%%%%%%%%%%%%%%%%%%%%%%%%%%%%%%%%%%%%%
%\section{Discussion}
\section{Conclusions}

We have carried out the calculations using the Skyrme SLy4 EDF
with and without the Coulomb EDF for nuclei with $Z=2-60$,
to find out the electromagnetic effect in nuclei.
In this paper, we focus our discussion on the quadrupole deformation and
the particle binding.
In many deformed nuclei,
the quadrupole deformation is enhanced by the Coulomb interaction.
On the other hand, the spherical nuclei stay spherical regardless of
the presence of the Coulomb interaction among protons.
It confirms that the quantum shell effects determine
whether the nucleus is deformed or not.

The Coulomb interaction affects not only the position of the proton drip line,
but also that of the neutron drip line, especially for nuclei with $Z\gtrsim 40$.
The neutron drip line moves to larger neutron numbers
and the region of the neutron-rich nuclei is extended.
This is unexpected because
the Coulomb interaction in the nucleus is repulsive and acts only among protons.
It seems to be due to a quantum mechanical effect:
The repulsive interaction makes the nuclear size larger,
which leads to a decrease in the kinetic energy of neutrons near the threshold.
An investigation in the canonical single-particle energies reveals that
the neutron energies near the threshold decrease
by the inclusion of the Coulomb interaction.

In this paper, we treat the exchange term of the Coulomb in the Slater approximation
\cite{VB72}.
However, we think that
the calculation of the exact Coulomb exchange term does not affect the conclusion,
because the magnitude of the exchange is significantly smaller than that of the direct term
and the Coulomb effect discussed in the present paper
is not directly associated with the exchange term.
We expect that the electromagnetic effects are larger
in the heavier systems, simply because
the proton number $Z$ and the Coulomb energy are larger.
It is desired to extend the present study to expand the region of the investigation
to rare-earth, actinide, and super-heavy nuclei.
Studies in this direction are currently underway.

\vspace{6pt}

\funding{
This work is supported by JST ERATO Grant No. JPMJER2304,
and KAKENHI Grant No. JP23K25864 and No. JP25K07312, Japan.
}
\begin{adjustwidth}{-\extralength}{0cm}

\reftitle{References}


\bibliography{main_rev1}

% If authors have biography, please use the format below
%\section*{Short Biography of Authors}
%\bio
%{\raisebox{-0.35cm}{\includegraphics[width=3.5cm,height=5.3cm,clip,keepaspectratio]{Definitions/author1.pdf}}}
%{\textbf{Firstname Lastname} Biography of first author}
%
%\bio
%{\raisebox{-0.35cm}{\includegraphics[width=3.5cm,height=5.3cm,clip,keepaspectratio]{Definitions/author2.jpg}}}
%{\textbf{Firstname Lastname} Biography of second author}

% For the MDPI journals use author-date citation, please follow the formatting guidelines on http://www.mdpi.com/authors/references
% To cite two works by the same author: \citeauthor{ref-journal-1a} (\citeyear{ref-journal-1a}, \citeyear{ref-journal-1b}). This produces: Whittaker (1967, 1975)
% To cite two works by the same author with specific pages: \citeauthor{ref-journal-3a} (\citeyear{ref-journal-3a}, p. 328; \citeyear{ref-journal-3b}, p.475). This produces: Wong (1999, p. 328; 2000, p. 475)

%%%%%%%%%%%%%%%%%%%%%%%%%%%%%%%%%%%%%%%%%%
%% for journal Sci
%\reviewreports{\\
%Reviewer 1 comments and authors’ response\\
%Reviewer 2 comments and authors’ response\\
%Reviewer 3 comments and authors’ response
%}
%%%%%%%%%%%%%%%%%%%%%%%%%%%%%%%%%%%%%%%%%%
% \PublishersNote{}
%\isPreprints{}{% This command is only used for ``preprints''.
\end{adjustwidth}
\end{document}